\documentclass[12pt]{article}

\usepackage{newtxtext,newtxmath}
\usepackage{graphicx}

\usepackage[letterpaper,margin=1in]{geometry}

\renewenvironment{abstract}
	{\quotation}
	{\endquotation}

\date{}

\makeatletter
\renewcommand{\fnum@figure}{\textbf{Figure \thefigure}}
\renewcommand{\fnum@table}{\textbf{Table \thetable}}
\makeatother

\usepackage{scicite}

\usepackage{url}

\usepackage[latin1]{inputenc}
\usepackage{amsmath}

\usepackage{amssymb}
\usepackage{graphicx,color}
\usepackage[dvipsnames]{xcolor}
\usepackage{braket,dsfont}
\usepackage{bm}
\usepackage{soul}
\usepackage[breaklinks]{hyperref}
\usepackage{verbatim}
\usepackage{parskip}
\usepackage{subfig}
\usepackage{comment}
\usepackage{upgreek}
\usepackage{float}
\usepackage{graphicx}
\usepackage{wrapfig}
\usepackage{lipsum}
\usepackage{afterpage}
\usepackage{etoolbox}
\usepackage{siunitx}
\usepackage{esint}
\usepackage{mathtools}
\usepackage{cleveref}
\usepackage{caption}
\usepackage{upgreek}

\newcommand{\pdf}{f}
\newcommand{\Om}{\Omega_\text{mix}}

\DeclarePairedDelimiter\floor{\lfloor}{\rfloor}
\crefname{equation}{}{}

\def\scititle{
	Tensor networks enable the calculation of turbulence probability distributions
}
\title{\bfseries \boldmath \scititle}

\author{
    Nikita Gourianov$^{1,2,\ast}$,
    Peyman Givi$^{1}$,
    Dieter Jaksch$^{2,3}$,
    and Stephen B. Pope$^{4}$\and
    \small{$^{1}$Department of Mechanical Engineering and Materials Science, University of Pittsburgh, Pittsburgh, PA 15261, USA}\and
    \small{$^{2}$Clarendon Laboratory, University of Oxford, Oxford OX13PU, UK}\and
    \small{$^{3}$Institut f{\"u}r Quantenphysik, Universit{\"a}t Hamburg, Luruper Chaussee 149, 22761 Hamburg, Germany}\and
    \small{$^{4}$Sibley School of Mechanical and Aerospace Engineering, Cornell University, Ithaca, NY 14853, USA}\and
    \small{$^\ast$Corresponding author. E-mail:  nikgourianov@icloud.com}
}

\begin{document} 

% Insert the title and author list
\maketitle

% Abstract, in bold
% There are strict length limits, and not all formats have abstracts.
% Consult the journal instructions to authors for details.
% Do not cite any references in the abstract.
\begin{abstract} \bfseries \boldmath
Predicting the dynamics of turbulent fluids has been an elusive goal for centuries. Even with modern computers, anything beyond the simplest turbulent flows are too chaotic and multi-scaled to be \textit{directly} simulatable. An alternative is to treat turbulence $\textit{probabilistically}$, viewing flow properties as random variables distributed according to joint probability density functions (PDFs). Such PDFs are neither chaotic nor multi-scale, yet remain challenging to simulate due to their high dimensionality. Here we overcome the dimensionality problem by encoding turbulence PDFs as highly compressed ``tensor networks'' (TNs). This enables single CPU core simulations that would otherwise be impractical even with supercomputers: for a $5+1$ dimensional PDF of a chemically reactive turbulent flow, we achieve reductions in memory and computational costs by factors of $\mathcal{O}(10^6)$ and $\mathcal{O}(10^3)$, respectively, compared to standard finite-difference algorithms. A future path is opened towards something heretofore thought infeasible: directly simulating high-dimensional PDFs of both turbulent flows, and other chaotic systems that can usefully be described probabilistically.
\end{abstract}

\section*{Introduction}

Despite the simple and deterministic physical laws governing it, turbulence remains an inherently complex and chaotic phenomenon. It is characterised by large numbers of eddies interacting in intricate and nonlinear ways across wide ranges of spatial and temporal scales, leading to the emergence of chaos. Making matters worse, practically important turbulent flows (e.g. fuel-oxidiser mixtures in combustion) often involve multiple chemically-reacting species, which introduces additional nonlinearities and scales. The presence of chaos prohibits predicting the exact dynamics of turbulent flow-fields over long periods of time, while the multi-scaled nature of the flow-fields makes their simulation immensely expensive due to the need to solve sets of coupled partial differential equations (PDEs) on very fine grids.

However, for practical applications it is rarely necessary to know the precise state of a turbulent flow-field at every point in space-time. Rather, one is typically more interested in far slower-varying \emph{statistical} quantities where the fluctuations are averaged out (such as the lift and drag of an aeroplane, or the rate of product formation in a chemical process). In the statistical description of turbulence, variables like velocities $\mathbf{U}$, chemical mass-fractions $\Phi_\alpha$, temperatures, etc., are treated as random variables (RVs) distributed according to some one-point, one-time joint probability density function (PDF) \cite{Hopf1952}
\begin{equation}
\pdf = \pdf(\mathbf{u}, \varphi_1,\ldots; \mathbf{x},t),
\end{equation}
across space $\mathbf{x}$ and time $t$, with $\mathbf{u}, \varphi_1$ being sample-space variables corresponding to $\mathbf{U}, \Phi_1$. The trajectory of $\pdf$ completely describes the one-point, one-time statistics of the flow dynamics~\cite{Monin2007}; which are the central quantities of interest in practical engineering calculations.

The time-evolution of $\pdf$ is modelled by Fokker-Planck PDEs that are straightforward to derive~\cite{Pope00,Fox03,Dopazo94}, but hard to solve. If $\pdf$ is $d$-dimensional, assigning $M$ points for each dimension results in a total of $M^d$ gridpoints. Given that $d$ can be as high as $\mathcal{O}(10^3)$ in realistic flows~\cite{Williams85,LNBG20}, direct schemes like finite-differences (FD) or volumes were long ago dismissed as computationally infeasible~\cite{Pope85} due to their seemingly exponential cost in $d$. This spurred the creation of indirect Monte Carlo (MC) algorithms for probabilistic turbulence simulations~\cite{Pope85}. These schemes have proven highly successful, enabling advanced turbulent combustion simulations involving thousands of CPU cores~\cite{Hiremath2012,Zhou2025}.
However, the randomness and slow convergence characterising MC methods can be avoided by directly solving the underlying Fokker-Planck equations.

It is not just probabilistic turbulence calculations that are hindered by the curse of dimensionality: quantum many-body systems are described by states whose sizes also grow exponentially (in the number of particles). However, \emph{physically-relevant} quantum states are known to be highly structured~\cite{Poulin2011}. Such structure can be exploited to compress the states into approximate, but highly accurate, polynomially large representations known as tensor networks (TNs). TN algorithms allow efficiently evolving these states and analyzing their physical properties without ever leaving the compressed TN representation~\cite{White1992, Vidal2003, Schollwock2011, Orus2019}, and have enabled the simulation of otherwise intractable quantum systems like superconductors, ferromagnets and quantum computers~\cite{Clark2004,Feiguin2007,Cheneau2012,Trotzky2012,Zheng2017, Huang2021,Zhou2020,Tindall2024, Begusic2024}. Recently, the TN formalism has begun spreading beyond quantum physics~\cite{Gourianov2022, Ye2022,Ye2024, Peddinti2024, Holscher2024, Ritter2023,Rohshap2024}.

Decades of empirical experience indicates that $\pdf$ is also highly structured. For instance, in homogeneous turbulence, velocities $\mathbf{U}$ are often distributed normally~\cite{Monin2007}, whereas mass-fractions $\Phi_\alpha$ have been observed to follow normal, exponential and beta distributions in non-reactive flows~\cite{JMMG96ali}. In more complicated reacting flows, the PDFs generally cannot be so simply parameterised~\cite{chen11}, although they remain smoother and more predictable than the underlying flow-fields \cite{NNGLP2017}.

This work shows that the structure contained in turbulence PDFs is readily exploitable through TNs: using a simple TN known as the ``matrix product state" (MPS) ansatz to encode $\pdf$ in a highly compressed format allows us to formulate a scheme for cheaply and directly solving the governing Fokker-Planck equations. When the PDF structure is well-matched to the MPS ansatz, the time-evolution costs just $\sim d \log M$; while standard FD schemes scale as $\sim M^d$. We demonstrate the advantage by looking at the following turbulent flow. %The TN formalism opens a path towards something heretofore regarded as infeasible: directly simulating the trajectories of high-dimensional PDFs of complex physical systems.

\section*{Results}

\subsection*{Probabilistic modelling of reactive turbulence}

\begin{figure}
\includegraphics[width=1.0\textwidth]{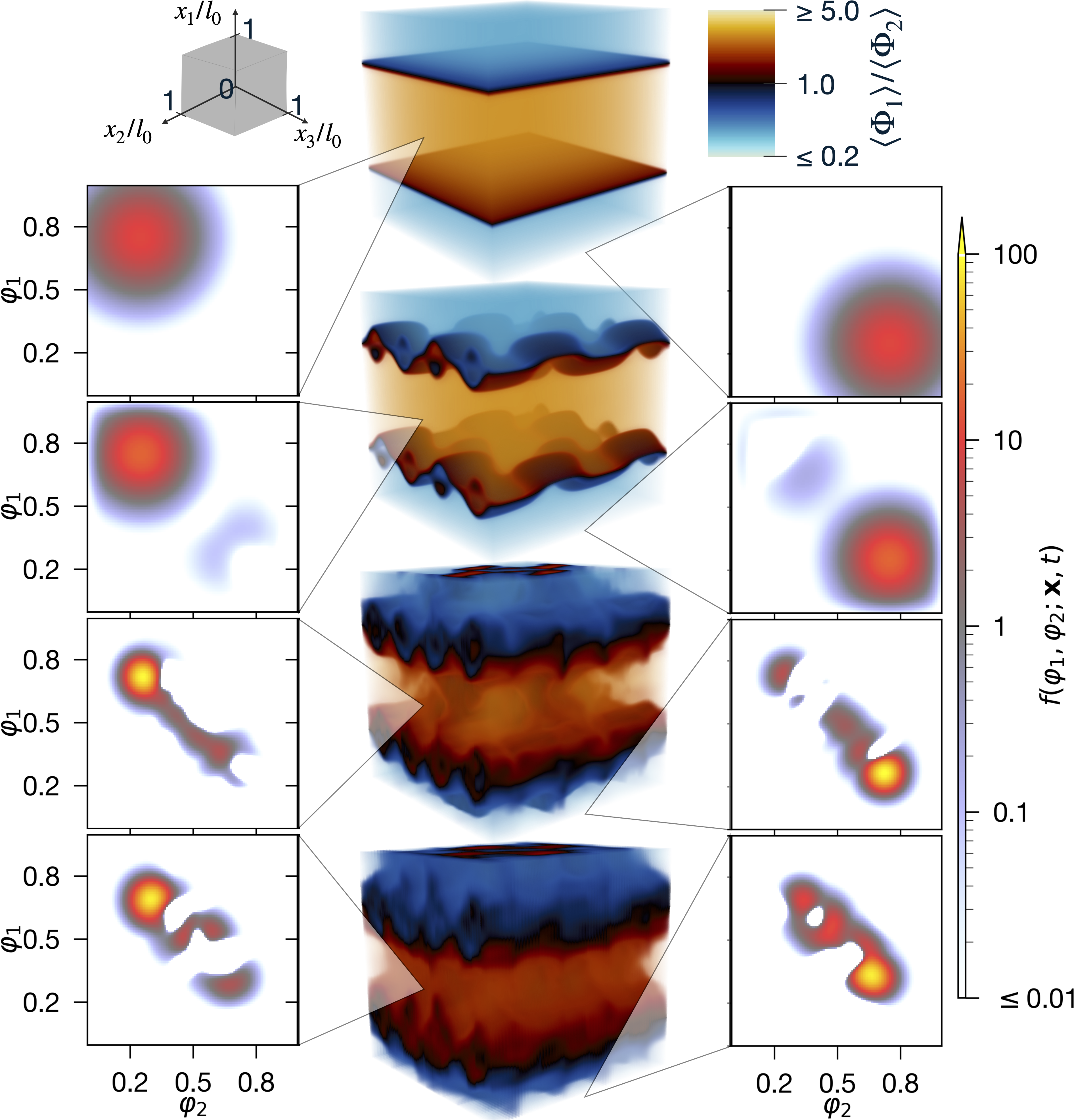}
\caption{\label{fig:1} \textbf{High-dimensional PDF of a flow undergoing turbulent mixing revealed by TN simulation}. Here the Fokker-Planck Eq.~\eqref{eq:FP0} is solved for a PDF $\pdf(\varphi_1,\varphi_2;\mathbf{x},t)$ over chemical mass-fractions $\varphi_1,\varphi_2$, at $C_\Omega = 1$, $\text{Da} = 0$ in the presence of a $\text{Pe} = 10^3$ velocity field $\braket{\mathbf{U}}$ characterised by vortices and a jet along $x_1$ ({Materials and Methods, Flow case definition}). The 5-dimensional $\pdf(\varphi_1,\varphi_2;\mathbf{x},t)$ is represented by a MPS ansatz at $\chi=128$, on a $128^{\times 5}$ grid and is visualised here for $\mathbf{x}/l_0 =(\frac{1}{2},1,\frac{1}{2})$ and $\mathbf{x}/l_0 = (0,\frac{1}{2},1)$ at times $t/T_0 = 0, 0.125, 1, 2$ in the left and right columns, while corresponding mean mass-fraction ratios $\braket{\Phi_1}/\braket{\Phi_2}$ are shown in the centre. }
\end{figure}

Consider an incompressible, 3D turbulent flow in which two chemical species are irreversibly reacting: $A + B \rightarrow \text{Products}$. In this system, two chemical species (of mass-fractions $\Phi_1,\Phi_2$) are stirred by a velocity field $\mathbf{U} = \mathbf{U}(\mathbf{x},t)$ in 3D space $\mathbf{x}$ across time  $t$. For the sake of simplicity, we here consider only the statistics of the $\Phi_1,\Phi_2$ scalar fields by assuming that the large-scale statistical features of the hydrodynamics are known a priori, while modelling the subgrid-scale (SGS), turbulent velocity fluctuations using large eddy simulation (LES), per current best practices~\cite{Zhou2025}. Doing so eliminates the randomness of $\mathbf{U}$ and reduces the dimensionality to $d=5+1$. Now, $\pdf = \pdf(\varphi_1, \varphi_2; \mathbf{x},t)$ describes the statistics of the mass-fraction \emph{fluctuations}, which provide the \emph{mean} mass-fractions $\braket{\Phi_1}, \braket{\Phi_2}$ through
\begin{equation}
\braket{\Phi_\alpha}(\mathbf{x},t) = \int_{[0,1)^{\times2}} \varphi_\alpha f(\varphi_1,\varphi_2 ;\mathbf{x},t) \text{d} \varphi_1 \text{d} \varphi_2.
\end{equation}
Such PDFs are known as ``filtered density functions''~\cite{Pope00,Givi06}. Deriving the equation governing $\pdf$ requires SGS closure modelling. Using popular closure models~\cite{Givi06} gives the Fokker-Planck PDE: 
\begin{equation}\label{eq:FP0}
\frac{\partial \pdf}{\partial t} + \braket{U_i} \frac{\partial \pdf}{\partial x_i} - \frac{\partial}{\partial x_i} \left[ (\gamma + \gamma_\text{SGS}) \frac{\partial \pdf}{\partial x_i} \right] = \frac{\partial}{\partial \varphi_\alpha} \left[ \Om \left( \varphi_\alpha - \braket{\Phi_\alpha} \right) \pdf \right]- \frac{\partial}{\partial \varphi_\alpha} ( S_\alpha \pdf).
\end{equation}
Here $\braket{\mathbf{U}}(\mathbf{x},t)$ is the large-scale (or, ``filtered") mean hydrodynamic field across $\mathbf{x} \in [0,l_0)^{\times3}$, $t\in[0,2 T_0)$, which is set to be a jet flow combined with a Taylor-Green vortex of amplitude $u_0 = l_0/T_0$ ({Materials and Methods, Flow case definition}).

The left hand side of Eq.~\eqref{eq:FP0} denotes the PDF-transport in space and time. The first term is the rate of temporal change, and the second term represents convection by the mean velocity field. The third represents the influence of the molecular ($\gamma$) and SGS diffusion [$\gamma_\text{SGS}(\mathbf{x},t)$] coefficients: the former sets the Peclet number $\text{Pe} = u_0 l_0/\gamma$, and the latter is modelled via the Smagorinsky \cite{Smagorinsky63} closure $\gamma_\text{SGS} = C_s \frac{\Delta_\ell^2}{2}  \sqrt{ \sum_{ij}\left(\frac{\partial U_i}{\partial x_j} + \frac{\partial U_j}{\partial x_i} \right)^2}$, with $C_s$ an empirical constant and $\Delta_\ell$ the LES filter width (both are specified in {Materials and Methods, Flow case definition}). 

The right hand side of Eq.~\eqref{eq:FP0} designates transport in the composition space  [``composition" since the $\varphi_1,\varphi_2 \in [0,1)^{\times2}$ mass-fractions define the composition of the fluid]. The first term represents scalar mixing from the SGS turbulence, and is modelled via the popular least-mean-square estimation (LMSE)~\cite{OBrien80} closure $\Om = C_\Omega \frac{\gamma + \gamma_\text{SGS}}{\Delta_\ell^2}$, with $C_\Omega$ the SGS mixing-rate. The final term denotes the effects of chemical reaction. For the binary reaction scheme considered here, $S_1 = S_2 = -C_r \varphi_1 \varphi_2$, where $C_r$ denotes the reaction rate that defines the Damk{\"o}hler number $\text{Da} = C_r l_0/u_0$. 

To solve Eq.~\eqref{eq:FP0}, we discretise $\pdf$ at every point in time on a $M=128,d=5$ Cartesian grid, but parameterise it as an MPS-network using far fewer variables than the $128^5$ gridpoints resolving it. This allows us to use a simple Runge-Kutta 2, FD scheme ({Materials and Methods, Finite-difference discretisation}) to solve Eq.~\eqref{eq:FP0} and time-evolve the MPS-PDF.

An initial MPS simulation ({Materials and Methods, Matrix product state algorithm}) is performed in Fig.~\ref{fig:1} of a purely mixing flow without chemical reactions ($\text{Da}=0$). The PDF is illustrated at two points in $\mathbf{x}$ along with the scalar-ratio $\braket{\Phi_1}/\braket{\Phi_2}$ at four different times, showing how the initially orderly, unmixed flow state is driven towards a fully-mixed $\braket{\Phi_1}/\braket{\Phi_2} \approx 1$ state by SGS and large-scale convective and diffusive mixing. The SGS mixing leads to the PDF concentrating, whilst the diffusion and mean-flow convection induces multi-modality in the PDF. The MPS simulation is highly accurate (Fig.~\ref{fig:2}b), yet the number of variables parameterising the PDF (NVPP) is only $\mathcal{O}(1/10^5)$ of an equivalent, classically-implemented FD scheme [{Materials and Methods, Eq.~\eqref{eq:MPSdofs}}].

\subsection*{Matrix product state encoding}

In our MPS encoding, the discretised, high-dimensional $\pdf(\varphi_1,\varphi_2,x_1,x_2,x_3)$ is decomposed into a 1D chain of tensors, where the $\varphi_1,\varphi_2,x_1,x_2,x_3$ dimensions are sequentially mapped to tensors from left to right, with each dimension itself decomposed into multiple tensors lengthscale-by-lengthscale [analogously to the ``sequential, serial" ordering in~\cite[Fig.~1]{Ye2024} and~\cite[Eq.~9]{Kiffner2023}]. This encoding exploits two separate structures that characterise the solution of Eq.~\eqref{eq:FP0}: firstly, the general smoothness of turbulence PDFs; secondly, that the different dimensions of $\pdf(\varphi_1,\varphi_2;\mathbf{x},t)$ are unlikely to be strongly coupled at low $C_\Omega$, because for $C_\Omega=0$ the PDF is separable ({Supplementary Text, Separability of Fokker-Planck equation}).

Matching the structure of the PDF in this way allows for an MPS encoding that is both accurate and parsimonious. The MPS representation (like any TN) can be systematically compressed, ie the NVPP reduced, by varying a hyperparameter known as the \emph{maximum bond-dimension} $\chi$. This hyperparameter regulates the maximal size of the ``bonds" between the tensors; which is equivalent to the maximum allowed coupling between the tensors and, in turn, between the different lengthscales and dimensions of $\pdf$. For example, setting $\chi=1$ forbids any coupling between the tensors and makes the NVPP minimal, while picking $\chi$ sufficiently large makes the MPS representation exact and $\text{NVPP} = M^5$ like in the standard representation. Setting $\chi$ to be small in turn leads to a low NVPP, but the MPS encoding will still remain accurate if it reflects the structure of $\pdf$ sufficiently well.

\subsection*{Validation of algorithm}

\begin{figure}\begin{center}
\includegraphics[width=1.0\textwidth]{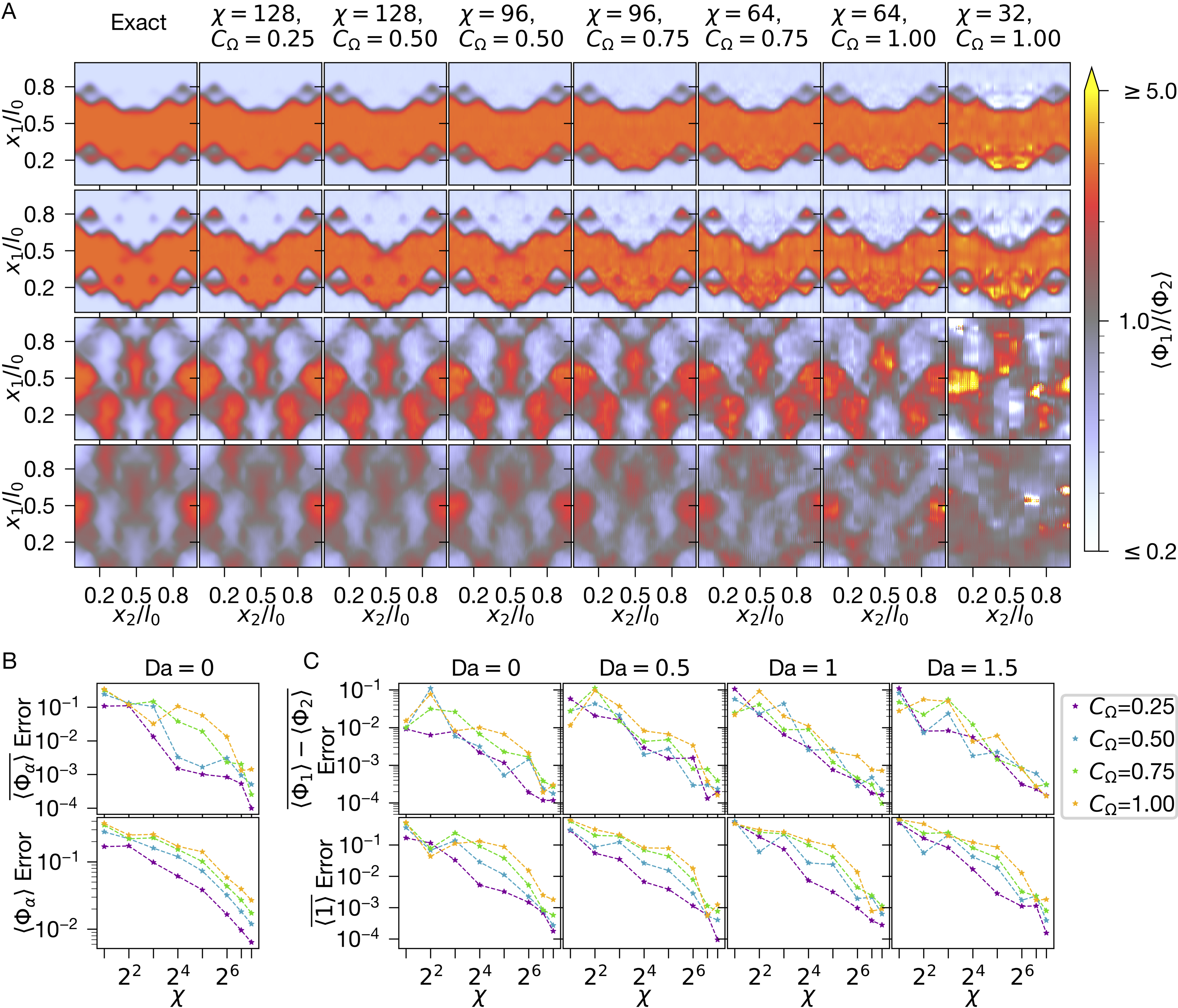}
\caption{\label{fig:2}
\textbf{Accuracy convergence of TN algorithm}. The influence of $\chi, C_\Omega$ and Da on the accuracy of the MPS simulation is outlined here. Panels \textbf{A} and $\textbf{B}$ contrast numerically exact means against those extracted from the MPS algorithm. In \textbf{A}, the ratio $\braket{\Phi_1}/\braket{\Phi_2}$ is visualised at $x_3/l_0=\frac{1}{2}$ for times $t/T_0 = 0.125, 0.25, 1, 2$, top to bottom. The leftmost column corresponds to the exact solution [which can only be practically computed for $\text{Da}=0$ through Eq.~\eqref{eq:FPMoms}], while the next six columns come from MPS simulations at varying $\chi, C_\Omega$. The differences between the exact and MPS solutions are quantified in the lower \textbf{B} plot; the upper \textbf{B} plot shows how well the total species amounts $\overline{\braket{\Phi_\alpha}}$ [see Eq.~\eqref{eq:spaceAvg}] are preserved through the simulation. In \textbf{C}, the RMSE in two basic statistics is computed: the difference in species consumption $\overline{\braket{\Phi_1} - \braket{\Phi_2}}$, which should always equal zero, and the space-averaged norm $\overline{\langle 1 \rangle}$, which should always equal one. All RMSEs are mathematically defined in {Materials and Methods, Error measures}.}
\end{center}
\end{figure}

We now investigate how well the MPS parameterisation fits the solution of Eq.~\eqref{eq:FP0} in practice. To determine the $\chi$ required to accurately simulate the dynamics of the RVs $\Phi_1, \Phi_2$ underlying the PDF, the composition space transport parameters $C_\Omega, \text{Da}$ are varied, while fixing the hydrodynamic variables $\braket{\mathbf{U}}$, $\gamma_\text{SGS}$ and $\text{Pe}$ to those used in Fig.~\ref{fig:1}.

Increasing $C_\Omega$ is expected to lead to higher coupling between the different dimensions of $\pdf$, which reduces the efficiency of our MPS encoding; i.e., increasing $C_\Omega$ requires an increased $\chi$ to maintain accuracy. To verify, we first set $\text{Da} = 0$ because this allows us to accurately compute $\braket{\Phi_\alpha}$ independently of Eq.~\eqref{eq:FP0} (see {Materials and Methods, Moment equations}) and benchmark the accuracy of the computed MPS-PDF across $C_\Omega, \chi$. The benchmark is shown in Figs.~\ref{fig:2}a and~\ref{fig:2}b. The $\braket{\Phi_1}/\braket{\Phi_2}$ ratios in Fig.~\ref{fig:2}a depict how the MPS-PDF means approach their numerically exact equivalent when $\chi$ increases and $C_\Omega$ decreases. All the cases, including the ones with lowest accuracy, correctly trend towards a fully-mixed equilibrium state where $\braket{\Phi_1}/\braket{\Phi_2} \approx 1$. Figure~\ref{fig:2}b quantitatively shows that the root-mean-square-error (RMSE) in terms of both the Reynolds-averaged mean mass-fraction
\begin{equation}\label{eq:spaceAvg}
\overline{\braket{\Phi_\alpha}} = \int_{[0,l_0)^{\times3}} \braket{\Phi_\alpha} \text{d} \mathbf{x}
\end{equation}
and $\braket{\Phi_\alpha}$ decrease roughly polynomially in $\chi$ for all $C_\Omega$.

Figure~\ref{fig:2}c depicts how varying the Damk{\"o}hler number affects the accuracy of the MPS algorithm. When $\text{Da}>0$, any moments $\braket{\Phi_\alpha^n},n\in \mathbb{Z}_{\geq0}$ higher than the norm $\braket{\Phi_\alpha^0} = \braket{1}$ can no longer be independently computed. We therefore rather look at two quantities which our simulation must preserve: the norm, which must equal unity across $\alpha,\mathbf{x},t$, and the difference in consumption between the two species $\overline{\braket{\Phi_1} - \braket{\Phi_2}}$, which should be zero for all $t$ due to the symmetry of $S_\alpha$ and the initial conditions. The figure indicates these two quantities becoming increasingly preserved when, again, $C_\Omega$ decreases and $\chi$ increases. Notably, the errors decrease roughly polynomially in $\chi$. However, varying $\text{Da}$ has little impact on the accuracy. This is because the chemical reaction largely just drives the PDF in compositional space towards the origin (as seen in Fig.~\ref{fig:3}), without significantly affecting its structure.

\subsection*{Computational complexity}

The maximal bond-dimension $\chi$ not only sets the accuracy of the MPS simulation, but also determines the computational cost. Because our 
MPS algorithm ({Materials and Methods, Matrix product state algorithm}) implements a finite-difference method within the MPS framework, it must perform the MPS-equivalent of operations like element-wise, matrix-matrix and matrix-vector multiplications, matrix and vector additions and subtractions, and inner and outer products, in addition to MPS-specific operations like singular values and QR decompositions to enforce the maximal bond-dimension and ensure the MPS stays in the numerically manageable ``canonical form"~\cite{Schollwock2011,Orus2019}. 
It is straightforward to show~\cite{Gourianov2022A} these MPS operations all cost $O( \chi^q d\log M)$ asymptotically, with 
$q\in \mathbb{Z}_{>0}$ depending on the operation.

The element-wise multiplication operation is the most expensive at $q=4$~\cite[Section 4.6]{Gourianov2022A}, making the asymptotic complexity of our scheme as a whole $O(\chi^4 d\log M)$ per timestep. Thus for $M_t$ timesteps, the total cost of the time-evolution will approach $O(M_t \chi^4 d\log M)$ at very large $\chi$; although in practice for small and intermediate $\chi$, the empirical cost scales much milder ({Supplementary Text, Empirical computational cost}). In comparison, standard FD schemes are exponentially more expensive in $d$, costing $O(M_t M^d)$.

There is also the question of preparing initial states and extracting statistics. Regarding the former, the 3D $\braket{\mathbf{U}}(\mathbf{x}),\gamma_\text{SGS}(\mathbf{x})$ and $d$-dimensional $\pdf(\varphi_1,\varphi_2; \mathbf{x}, t=0)$ can be computed using either the prolongation method [see~\cite[Sec.~4.4]{Gourianov2022A} and~\cite{Lubasch2018}] or the tensor-cross algorithm~\cite{Oseledets2010,Dolgov2020, Ritter2023, Ghahremani2024,Fernandez2024}, both at $O( \chi^3 d \log M)$ cost. As for the latter, computing expectation values boils down to doing the MPS-equivalent of matrix-vector multiplication and inner products, which are, as noted previously, inexpensive and straightforward operations. For instance, at any given timestep, the 3D mean $\braket{\Phi_\alpha}$ can be extracted from $\pdf$ at $O(M^3 \chi^2 d\log M)$ complexity, while the cost is $O(\chi^2 d\log M)$ for the scalar $\overline{\braket{\Phi_\alpha}}$.

\subsection*{Integrated quantities}

\begin{figure}\begin{center}
\includegraphics[width=0.75\textwidth]{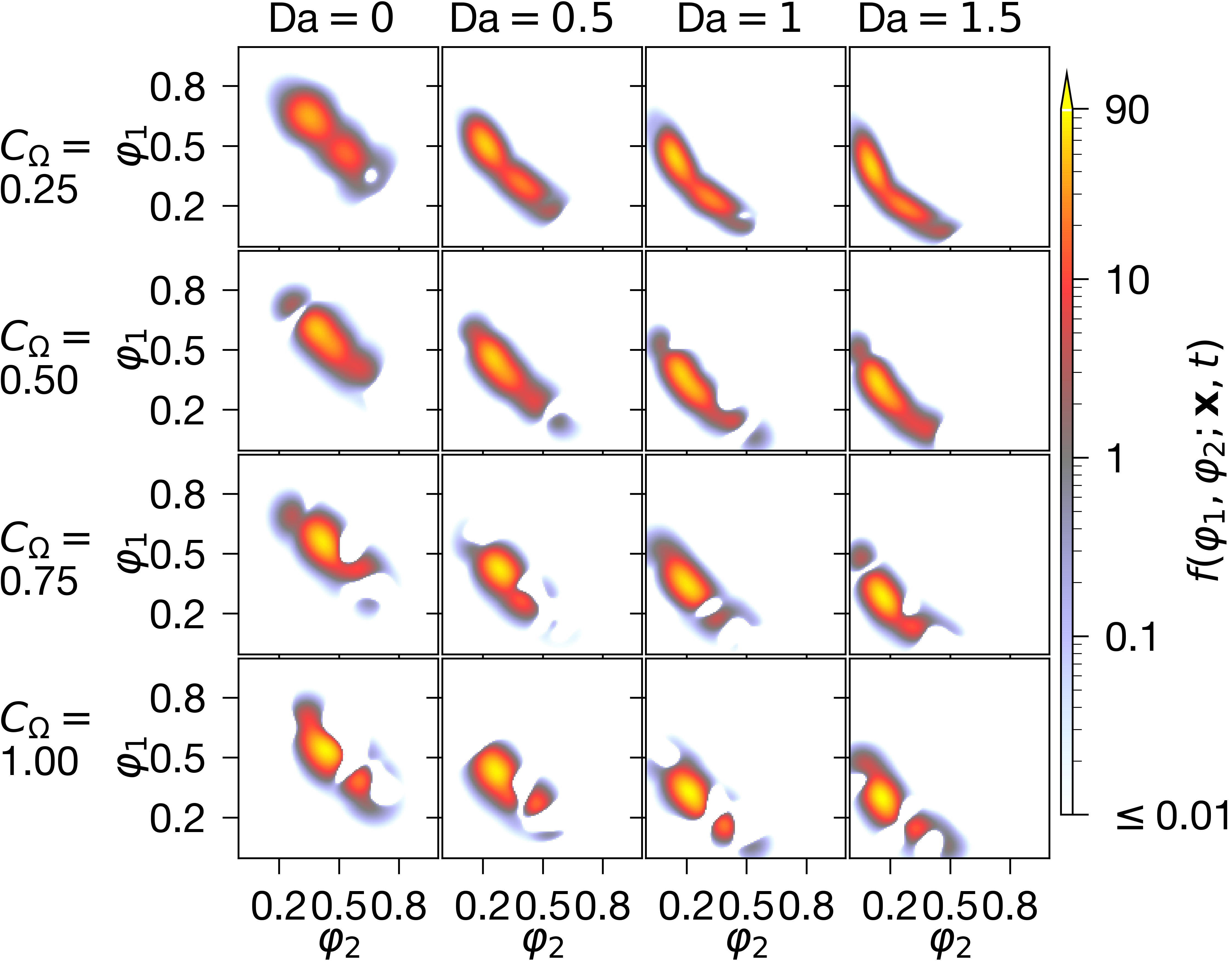}
\caption{\label{fig:3}
\textbf{Final PDF for various flow parameters}. The PDFs at the end of the simulation ($t/T_0=2$) are shown here in the centre of the spatial domain [$\mathbf{x}/l_0 = (\frac{1}{2},\frac{1}{2},\frac{1}{2})$] for all combinations of $C_\Omega, \text{Da}$. The PDFs are computed using $\chi = 128$ MPS simulations.}
\end{center}
\end{figure}

The satisfactory accuracy and subexponential cost of our MPS scheme allows us to directly compute the PDF, visualise it and extract from it all relevant integrated quantities. 

Figure~\ref{fig:3} shows the influence of mixing and chemical reactions on the PDF. As expected, in the absence of chemical reaction, both species tend toward the fully mixed values $\braket{\Phi_\alpha} (t\rightarrow \infty) \rightarrow 0.5$ at a rate governed by $C_\Omega$. Whereas in the reacting flow simulations, $\braket{\Phi_\alpha}(t\rightarrow \infty) \rightarrow 0$ at a rate that increases with $C_\Omega$ and $\text{Da}$. Visually, we see that increasing $C_\Omega$ leads to a PDF that is more concentrated along $\varphi_1=\varphi_2$ (implying a more mixed fluid), while increasing $\text{Da}$ takes the PDF closer to the origin (meaning more of the reactants have been consumed). Multi-modality is also evident in some PDFs; this is a result of convective and diffusive transport in $\mathbf{x}$-space.

\begin{figure}\begin{center}
\includegraphics[width=0.75\textwidth]{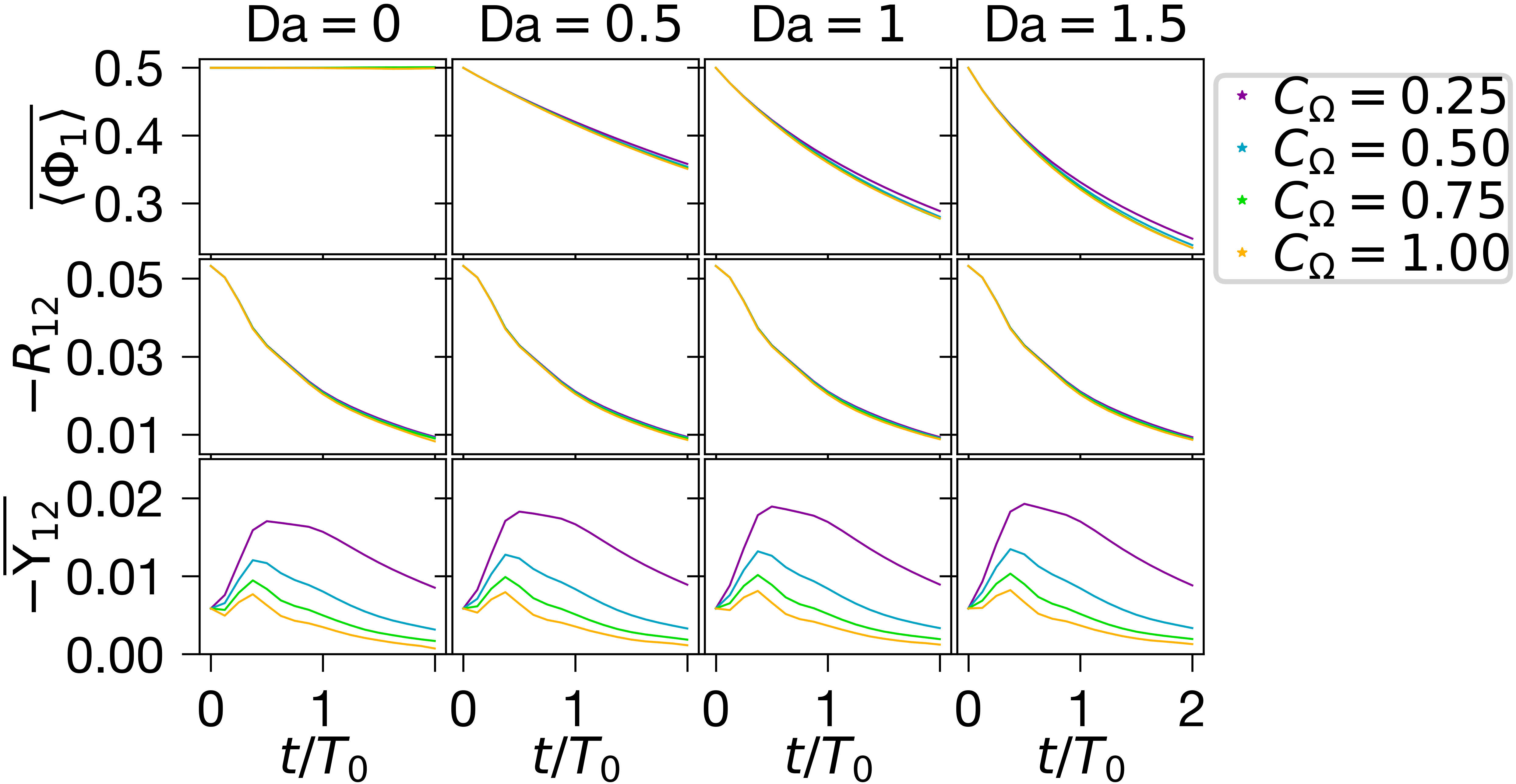}
\caption{\label{fig:4}
\textbf{Statistics extracted from PDF for different flow parameters}. In the first row, the total amount of the first species is plotted. The next row is of the Reynolds-averaged covariance $R_{12}$, while the final row quantifies the space-averaged  covariance $\overline{\text{Y}_{12}}$. These quantities are defined in Eqs.~\cref{eq:spaceAvg,eq:RACov,eq:SGSCov}. The PDFs are computed at $\chi=128$. }
\end{center}
\end{figure}

The trends of Fig.~\ref{fig:3} are reflected in the integrated quantities plotted in Fig.\ \ref{fig:4}. %There, the first and second moments are extracted from a PDF that is now \emph{normalised} pointwise in $\mathbf{x},t$ (to eliminate the norm-error plotted in Fig.~\ref{fig:2}c). 
The first row illustrates $\overline{\braket{\Phi_1}}$ going from being conserved at $\text{Da} = 0$, to consumed at rates increasing with $\text{Da}$, as expected. The consumption also slightly increases with the SGS mixing-rate.
The following row shows the negative of the (Reynolds averaged) scalar covariance
\begin{equation}\label{eq:RACov}
R_{12} = \overline{\braket{\Phi_1}\braket{\Phi_2}} - \overline{\braket{\Phi_1}}\text{ }\overline{\braket{\Phi_2}},
\end{equation}
which decays in $t$ due to $\braket{\Phi_1}/\braket{\Phi_2}$ approaching unity as the flow becomes increasingly mixed. Finally, the last row exhibits the covariance
\begin{equation}\label{eq:SGSCov}
\overline{\text{Y}_{12}} = \overline{\braket{\Phi_1 \Phi_2} - \braket{\Phi_1}\braket{\Phi_2}}.
\end{equation}
At initial times, $-\overline{\text{Y}_{12}}$ increases due to large gradients in $\braket{\Phi_\alpha}$, followed by a decrease due to mixing (with higher mixing-rates leading to a faster decay). The ratio $R_{12}/(R_{12}+\overline{\text{Y}_{12}})$ is consistently above one half, implying most of the energy of the eddies is resolved during the simulations. The statistical trends observed in Fig.~\ref{fig:4} are consistent with those reported in turbulence literature~\cite{Givi06}.

\section*{Discussion}
The results imply that MPSs are able to efficiently exploit structure within turbulence PDFs. The PDF $\pdf(\varphi_1,\varphi_2;\mathbf{x},t)$ of our 3D chemically-reactive flow case [Eq.~\eqref{eq:FP0}] is of an orderly shape, and the coupling between its dimensions is limited by the SGS mixing-rate $C_\Omega$. Exploiting these structures permits our MPS scheme to accurately and efficiently represent $\pdf$ and evolve it through time. In the future, a more realistic model should be considered where the velocity field is also included within the PDF, turning $\pdf = \pdf(\mathbf{u}, \varphi_1,\varphi_2;\mathbf{x},t)$ into a $d=8+1$ dimensional object.

Ensuring the MPS algorithm maintains both accuracy and efficiency requires carefully selecting $\chi$ (see Figs.~S1 to S3). Figure~\ref{fig:2} indicates that varying the Damk{\"o}hler number does not significantly affect the accuracy, while increasing the SGS mixing-rate requires $\chi$ to in turn increase as $\chi \sim \text{poly}(C_\Omega)$ for accuracy to be maintained. Setting $\chi$ excessively high is expensive due to the $O(\chi^4 d\log M)$ asymptotic cost of the algorithm. However, for slower mixing-rates, high accuracy is achievable at very low $\chi$ even when the chemical reaction rates are high. For instance, at $C_\Omega=0.25$, $\text{Da} = 1.5$, the algorithm is accurate with just $\chi=32$. This is equivalent to respective $\mathcal{O}(10^6)$ and $\mathcal{O}(10^3)$ factor reductions in memory and computational costs [{Materials and Methods, Eq.~\eqref{eq:MPSdofs}} and {Supplementary Text, Empirical computational cost}] compared to conventional FD schemes; allowing the time-evolution to be executed on a single CPU core in only a couple of hours, instead of days on a supercomputer.

The results shown here are only an early indication of what is possible: there exists great scope for improvement in both the algorithm and its implementation.
For example, employing tensor-cross or other algorithms~\cite{Michailidis2024} to perform element-wise multiplications might reduce the complexity of our scheme to $\sim \chi^3$ without significantly sacrificing accuracy. Furthermore, better optimised software running on specialised computing architectures will allow for much larger bond-dimensions and system sizes: we are currently simulating a $M^d = 128^5$ $=2^{35}$ grid at $\chi=128$, while the current record is a quantum physics simulation on a grid-equivalent of $M^d = 2^{400}$ at $\chi=32768$, performed on a tensor processing unit pod~\cite{Ganahl2023}.

We decided to use an MPS ansatz because it closely matches the structure of the PDF for this particular flow; other flows may have different PDFs for which alternative ansatze could be better suited~\cite{Glasser2019}. Fortunately, there exists a rich and growing selection of TNs to pick from, each carrying their advantages and disadvantages. These range from 2D generalisations of MPSs~\cite{Verstraete2006A}, hierarchical networks~\cite{Murg2010,Evenbly2014} and even networks that might some day leverage quantum hardware~\cite{Jaksch2023}. There is also the exciting prospect of catering the TN ansatz to the structure of the PDF in an automated manner~\cite{Peng2023}. Typically, more complex ansatze are able to encode solutions with higher accuracy at lower $\chi$, but are costlier to manipulate. Balancing such considerations while exploring alternative TN geometries for probabilistic turbulence simulations is a promising avenue of future investigation.

Turbulence is just one example of a complex system; there are many others, ranging from biological organisms to financial markets~\cite{Favre1995}. These kind of systems exhibit chaotic and unpredictable dynamics that ultimately require statistical descriptions~\cite{Bandak2024}. The most fundamental way of doing so is by modelling their PDFs. Yet, such PDFs are typically prohibitively high-dimensional (as displayed here for the case of turbulence) which has made solving their governing Fokker-Planck equations infeasible, until now. This work is a first demonstration in how the problem can be overcome via a simple TN. More advanced TN ansatze and algorithms will be developed in time, holding the promise of enabling large-scale probabilistic simulations both within the field of fluid dynamics, and beyond.

\section*{Materials and Methods}

\subsection*{Flow case definition}
In Eq.~\eqref{eq:FP0}, the mean velocity field $\braket{\mathbf{U}}$ is set a priori. To ensure adequate convective mixing and for the flow to be interesting, we elected to set the velocity field to a jet moving through a Taylor-Green vortex:
\begin{equation}\label{eq:velField}
\begin{split}
\braket{U_1}/u_0 &=  \cos{k x_1} \sin{k x_2} \sin{k x_3} - e^{- \frac{(x_2/l_0 -1/2)^2 + (x_3/l_0 - 1/2)^2 }{2 (1/6)^2}},\\
\braket{U_2}/u_0 &=  \sin{k x_1} \cos{k x_2} \sin{k x_3},\\
\braket{U_3}/u_0 &= -2\sin{k x_1} \sin{k x_2} \cos{k x_3}.
\end{split}
\end{equation}
Here, the vortex wavenumber $k$ is set to $k = 4 \pi/l_0$.

The initial ($t=0$) PDF is chosen to be a Gaussian step-function
\begin{equation}\label{eq:initPDF}
\pdf(t=0) = \frac{1}{{2 \pi} (1/8)^2} \left\{\begin{array}{ll} e^{-\frac{(\varphi_1 - 3/4)^2 + (\varphi_2 -1/4)^2}{2(1/8)^2}}, & \frac{1}{4} \leq x_1 < \frac{3}{4}, \\ e^{-\frac{(\varphi_1-1/4)^2 + (\varphi_2-3/4)^2}{2(1/8)^2}}, & \mbox{otherwise,}\end{array}\right.
\end{equation}

that has undergone numerical smoothing in the $\mathbf{x}$ dimensions (the smoothing is meant to soften the step-function at the $x_1 = \frac{1}{4}, \frac{3}{4}$ boundary sufficiently to avoid numerical instabilities during time-evolution). The initial PDF is illustrated in the first rows of Fig.~\ref{fig:1}.

The $M=128,d=5$ grid is sufficient for the simulation to be conducted with parameters that make physical sense: we set $C_s = 0.11$, $\Delta_\ell = 3 \Delta x = 3l_0/M$, $\text{Pe} = 10^3$, $C_\Omega \in [0.25,1]$ and $\text{Da} \in [0,1.5]$.

\subsection*{Moment equations}
The zeroth moment of the Fokker-Planck Eq.~\eqref{eq:FP0} recovers the hydrodynamic continuity equation $\nabla \cdot \braket{\mathbf{U}}=0$, while the first gives an equation for the mean mass-fractions $\braket{\Phi_{\alpha}}(\mathbf{x},t)$: 
\begin{equation}\label{eq:FPMoms}
\frac{\partial \braket{\Phi_{\alpha}}}{\partial t} + \braket{U_i}\frac{\braket{\Phi_{\alpha}}}{\partial x_i} = \frac{\partial}{\partial x_i}\left[ (\gamma + \gamma_\text{SGS})\frac{\partial \braket{\Phi_{\alpha}}}{\partial x_i} \right] + \braket{S_{\alpha}}.
\end{equation}
In non-reactive flows ($\text{Da}=0$), $S=0$ and Eq.~\eqref{eq:FPMoms} can be cheaply and accurately solved using a standard FD scheme to obtain a ``numerically exact" $\braket{\Phi_{\alpha}}$ solution (in the sense that there is no truncation error in $\chi$, as explained in {Materials and Methods, Error measures}). This is used to check the accuracy of the MPS algorithm in Figs.~\ref{fig:2}a and~\ref{fig:2}b. It is not possible to obtain a numerically exact $\braket{\Phi_{\alpha}}$ when $\text{Da}>0$, because a closure model would be required for $\braket{S_{\alpha}}$.

\subsection*{Finite-difference discretisation}
The simulations are performed on equidistant Cartesian grids with $M=128$ gridpoints along each dimension. The derivatives in Eqs.~\eqref{eq:FP0} and~\eqref{eq:FPMoms} are discretised in a simple manner: the temporal derivative with an explicit Runge-Kutta 2 scheme, and a second-order-accurate central finite-differences (CFD2) discretisation of the $\mathbf{x},\varphi_1,\varphi_2$ derivatives.

However, discretising Eq.~\eqref{eq:FP0} creates the practical challenge of handling delta-functions. The LMSE model forces each $\Phi_\alpha$ towards $\braket{\Phi_\alpha}$ at every $\mathbf{x},t$, equivalent to the PDF in composition space moving towards a delta-function centered around the mean of the mass-fractions. Resolving delta-functions on discretised grids is difficult, as their sharp gradients reduce the accuracy and stability of any numerical scheme used to compute the PDF-transport. While often this is dealt with by employing highly dissipative discretisations of derivatives (e.g. upwinding), we rather choose to simply modify the LMSE model in Eq.~\eqref{eq:FP0} through the addition of an \emph{artificial dissipation} term to the compositional space. Doing this while discretising the Fokker-Planck PDE results in:
\begin{equation}\label{eq:FP2}
\begin{split}
&\frac{\Delta \pdf}{\Delta t} + \braket{U_i} \frac{\Delta \pdf}{\Delta x_i} - \frac{\Delta}{\Delta x_i} \left[ (\gamma + \gamma_\text{SGS}) \frac{\Delta \pdf}{\Delta x_i} \right] = \\ 
&\frac{\Delta}{\Delta \varphi_\alpha} \left[ \Omega_\text{mix} \left( \varphi_\alpha - \braket{\Phi_\alpha} \right) \pdf + C_\Omega\mu \frac{\Delta f}{\Delta \varphi_\alpha} \right] - \frac{\Delta}{\Delta \varphi_\alpha} ( S_\alpha \pdf ),
\end{split}
\end{equation}
with the artifical dissipation governed by $\mu$. This parameter needs to be set to be as small as possible to minimally affect the accuracy, while still being large enough to ensure $\pdf$ is well-resolved on $M$. From trial and error, we find $\mu = 4\cdot 10^{-3} \frac{u_0}{l_0}$ works well for $M=128$. 

Particular care must be taken when defining the boundary conditions for this problem. While in $\mathbf{x}$-space one may simply assume periodic boundaries, in compositional space the boundary conditions must be defined in a way that stops probability leaking out of the domain. This is achieved by making the composition-space ghosts points for any order-$n$ discrete derivative of $\pdf$ follow
\begin{equation}\label{eq:compBCs}
\sum_{i = 0}^{M-1} \left[\frac{\Delta^n \pdf}{\Delta \varphi_\alpha^n}\right]_i = 0,
\end{equation}
with $i$ denoting a discretised (equidistantly distributed) lattice point. Equation~\eqref{eq:compBCs} imposes $\pdf_{-1} = -\pdf_0 \text{ \& } \pdf_M = -\pdf_{M-1}$ for the first derivative, and $\pdf_{-1} = \pdf_0 \text{ \& } \pdf_M = \pdf_{M-1}$ for the second, under our CFD2 discretisation.

\subsection*{Matrix product state algorithm}

Our MPS algorithm implements the aforementioned RK2-CFD2 scheme on the MPS manifold~\cite{Schollwock2011,Holtz2012}. This entails parameterising \emph{all} the vectors (like $\pdf$, $\braket{U_i}$ and $\braket{\Phi\alpha}$) in Eq.~\eqref{eq:FP2} as MPSs~\cite[Section 3.4]{Gourianov2022A}, and the matrices (e.g. $\Delta/\Delta x_i$) as analogous matrix product operators~\cite[Section 3.5]{Gourianov2022A}. Then, within the MPS format, the time-stepping is performed in a standard manner using the arithmetic operations outlined in the \emph{Computational complexity} section.

It is essential to control the bond-dimension during the MPS simulation. The arithmetic operations that time-evolve $\pdf$ lead to its bond-dimension growing exponentially in time, if not truncated~\cite[Section 3.5.3]{Gourianov2022A}. In our code, we employ the singular values decomposition to truncate the bond-dimension of $\pdf$ such that it is always limited to $\chi$. As for the other vectors and matrices, these objects remain constant in time and their bond-dimensions are all of order $\mathcal{O}(10)$.

The maximal bond-dimension $\chi$ defines the NVPP. For an MPS representation of $\pdf$, the number of parameters becomes
\begin{equation}\label{eq:MPSdofs}
\text{NVPP} = 2 \sum_{n=1} ^N p(n{-}1) p(n) - \sum_{n=1}^{N-1} p(n)^2,
\end{equation}
with $N = \log_2 M^d$, ($M$ must be a power of $2$) being the number of tensors in the MPS, and $p(n) = \min(2^n, 2^{N-n},\chi)$ being the size of the $n$th bond of the MPS. The first sum gives the total number of parameters in the MPS, while the second sum represents the intrinsic gauge degrees of freedom of the MPS format~\cite{Holtz2012}. When $\chi$ is maximal, i.e. $\chi=2^{\floor{N/2}}$, we get $\text{NVPP}=2^{N}=M^d$ and that $\pdf$ is represented \emph{exactly} on the $M^{\times d}$ grid.

\subsection*{Error measures}
The errors in Figs.~\ref{fig:2}b and~\ref{fig:2}c are computed using the RMSE measure across $\chi$. In the first figure, the upper $\text{Error}_{\text{2b}\uparrow}$ and lower $\text{Error}_{\text{2b}\downarrow}$ are computed by averaging the spatially-averaged mean-quantities across $\alpha,t$ and $\alpha,t,\mathbf{x}$, respectively. In Fig.~\ref{fig:2}c, the averaging is done across just $t$ and $t,\mathbf{x}$ to compute $\text{Error}_{\text{2c}\uparrow}$, $\text{Error}_{\text{2c}\downarrow}$. Mathematically, these errors can be expressed as:
\begin{equation}\label{eq:RMSE0}
\begin{split}
&\text{Error}_{\text{2b}\uparrow}(\chi) = \sqrt{\frac{1}{2} \sum_{\alpha=1,2} \mathcal{E}_t\Big[\overline{\braket{\Phi_\alpha}}(\chi), 1/2 \Big]},\\
&\text{Error}_{\text{2b}\downarrow}(\chi) = \sqrt{\frac{1}{2} \sum_{\alpha=1,2} \mathcal{E}_{t,\mathbf{x}}\big[\braket{\Phi_\alpha}(\chi),\braket{\Phi_\alpha}(\text{exact}) \big]},\\
&\text{Error}_{\text{2c}\uparrow}(\chi) = \sqrt{\mathcal{E}_t\Big[\overline{\braket{\Phi_1}}(\chi)-\overline{\braket{\Phi_2}}(\chi), 0 \Big]},\\
&\text{Error}_{\text{2c}\downarrow}(\chi) = \sqrt{\mathcal{E}_{t,\mathbf{x}}\big[\braket{1}(\chi),1 \big]},
\end{split}
\end{equation}
with $\braket{\Phi_\alpha}(\chi)$ being extracted from the MPS-PDF solution of Eq.~\eqref{eq:FP2} while $\braket{\Phi_\alpha}(\text{exact})$ is the solution found by directly solving Eq.~\eqref{eq:FPMoms} with a standard RK2-CFD2 FD scheme; this solution is ``numerically-exact" in the sense that it does not suffer from any truncation error in $\chi$ [although a truncation error from the FD discretisation itself remains, this error is slight due to the smoothness of $\braket{\Phi_\alpha}(\text{exact})$]. The $\mathcal{E}_t$ function implements temporal-averaging, while $\mathcal{E}_{t,\mathbf{x}}$ performs averaging across space-time:
\begin{equation}\label{eq:RMSE1}
\begin{split}
\mathcal{E}_{t,\mathbf{x}}\left(g,g_0 \right) =&\int_{[0,l_0)^{\times3}} \frac{\text{d}\mathbf{x}}{l_0^3} \mathcal{E}_t\left[g(\mathbf{x}),g_0(\mathbf{x}) \right],\\
\mathcal{E}_t\left(g,g_0 \right) =& \int_0^{2T_0} \frac{\text{d}t}{2T_0} \big[g(t)-g_0(t)\big]^2.
\end{split}
\end{equation}

Note that since both time and space are discretised during the simulations, the above integrals are both performed numerically using a simple step-quadrature. In space, the integrals are computed using all the $M = 128$ gridpoints along each dimension. In time, the integral is computed over the $17$ time-samples $t=0,\frac{T_0}{8},\frac{T_0}{4}, \frac{3T_0}{8}, \ldots, 2 T_0$.

%%%%%%%%%%%%%%%% MAIN TEXT FIGURES %%%%%%%%%%%%%%%

%%%%%%%%%%%%%%%% REFERENCES %%%%%%%%%%%%%%%

\clearpage % Clear all remaining figures and tables then start a new page

% The list of references goes after the main text and before the acknowledgements
% When preparing an initial submission, we recommend you use BibTeX, like this:
%

\bibliographystyle{sciencemag}

% After the paper has completed peer review and been revised ready for acceptance,
% you should comment out the lines above and copy-paste the contents of your .bbl
% file here instead. This will help ensure that our conversion software works correctly.
% Remember to re-run BibTeX first - check the timestamp!
%
% Example of the first three entries copy-pasted from science_template.bbl:
%
%\begin{thebibliography}{1}
%
%\bibitem{example}
%A.~N. {Author}, An example reference. \emph{Journal of Improbable Research}
%  \textbf{1}, 67 (2020).
%
%\bibitem{example2}
%F.~M. {Surname}, S.~{Author}, A second example. \emph{Interesting Research
%  Letters} \textbf{32}, 897 (2019).
%
%\bibitem{example_preprint}
%P.~{One}, P.~{Two}, P.~{Three}, {An unpublished preprint}. \emph{preprint}
%  (2021), arXiv:2101.12345.
%
%\end{thebibliography}

%%%%%%%%%%%%%%%% ACKNOWLEDGEMENTS %%%%%%%%%%%%%%%

\section*{Acknowledgments}
We thank Professor Ard Louis at the University of Oxford for fruitful discussions and would like to acknowledge the use of the University of Oxford Advanced Research Computing (ARC) facility in carrying out this work (\url{http://dx.doi.org/10.5281/zenodo.22558}).
\paragraph*{Funding:} NG acknowledges support by US-AFSOR grant FA8655-22-1-7027 and the UKRI ``Quantum Computing and Simulation Hub" grant EP/P009565/1.
PG acknowledges support by US-AFOSR grant FA9550-23-1-0014.
DJ acknowledges support by the European Union's
Horizon Programme (HORIZON-CL42021-DIGITALEMERGING-02-10) Grant Agreement 101080085 QCFD, the Cluster of Excellence ``Advanced Imaging of Matter"
of the Deutsche Forschungsgemeinschaft (DFG)- EXC 2056- project ID 390715994, and the Hamburg Quantum Computing Initiative (HQIC) project EFRE. The project is co-financed by ERDF of the European Union and by ``Fonds of the Hamburg Ministry of Science, Research, Equalities and Districts (BWFGB)".
\paragraph*{Author contributions:} N.G. and P.G. conceived and planned the research project. N.G. formulated the matrix product state algorithm, did the analytical calculations and wrote the software. The numerical experiments were jointly designed by all the authors and executed by N.G. The numerical results were analysed and interpreted jointly by all the authors. N.G. and P. G. wrote the manuscript, with contributions from the other authors. The project was supervised by P. G. and S. B. Pope.

\paragraph*{Competing interests:} There are no competing interests to declare.
\paragraph*{Data and materials availability:} All software and data accompanying this manuscript are publicly available at \url{https://github.com/nikitn2/tendeq} and \url{https://doi.org/10.5281/zenodo.14223424}.

%%%%%%%%%%%%%%%% SUPPLEMENT LIST %%%%%%%%%%%%%%%

% List the contents of your Supplementary Materials, including the numbers of any
% supplementary figures, tables, external data files etc. and any references that are
% cited only in the supplement. In this example, refs. 7-8 are cited only in the supplement.
% Fill out your numbers accordingly and delete any lines that aren't applicable.
\subsection*{Supplementary Materials}
Supplementary Text\\
Figs. S1 to S3\\
Reference \textit{(\arabic{enumiv})}\\ % automatically fills out the last reference number
% (filling out the other numbers automatically is possible but fiddly and liable to break)

%%%%%%%%%%%%%%%% END OF MAIN TEXT %%%%%%%%%%%%%%%

\newpage

%%%%%%%%%%%%%%%% START OF SUPPLEMENT %%%%%%%%%%%%%%%

% Figures, tables, equations and pages in the supplement are numbered S1, S2 etc.
\renewcommand{\thefigure}{S\arabic{figure}}
\renewcommand{\thetable}{S\arabic{table}}
\renewcommand{\theequation}{S\arabic{equation}}
\renewcommand{\thepage}{S\arabic{page}}
\setcounter{figure}{0}
\setcounter{table}{0}
\setcounter{equation}{0}
\setcounter{page}{1} % not 0 as \newpage already started a supplementary page
% References continue the numbering from the main text.

%%%%%%%%%%%%%%%% SUPPLEMENT TITLE PAGE %%%%%%%%%%%%%%%

\begin{center}
\section*{Supplementary Materials for\\ \scititle}
Nikita Gourianov$^{\ast}$,
Peyman Givi,
Dieter Jaksch,
and Stephen B. Pope\\
\small$^{\ast}$ Corresponding author. E-mail:  nikgourianov@icloud.com
\end{center}

% Fill out the numbers for each type of supplementary material,
% and delete any lines that aren't applicable.
% These are just example numbers that don't match the rest of this template.
\subsubsection*{This PDF file includes:}
Supplementary Text\\
Figures S1 to S3\\

\newpage

%%%%%%%%%%%%%%%% MATERIALS AND METHODS %%%%%%%%%%%%%%%

\subsection*{Supplementary Text}
\subsubsection*{Separability of Fokker-Planck equation}
Here we explain how the Main Text Fokker Planck Equation~(2) becomes separable when $C_\Omega = 0$. Assume the PDF can be written in the form 

\begin{equation}
f(\varphi_1,\varphi_2,x_1,x_2,x_3,t) = g(x_1,t) h(\varphi_1,t) R(\varphi_2,x_2,x_3,t).
\end{equation}
Inserting this into Equation~(2), gives
\begin{equation}\label{Eq:FP}
\begin{split}
&hR \left( 
\frac{\partial g}{\partial t} + \braket{U_1} \frac{\partial g}{\partial x_1} - \frac{\partial }{\partial x_1} \left[ \left( \gamma + \gamma_\text{SGS} \right) \frac{\partial g}{\partial x_1}\right] \right) +\\
&gR \left( \frac{\partial h}{\partial t}  - \Omega_\text{mix}\frac{\partial}{\partial \varphi_1}\left[ (\varphi_1 -\braket{\Phi_1} )h\right] - C_r \varphi_2\frac{\partial}{\partial \varphi_1} [\varphi_1 h]\right)+ \\
& gh \bigg( \frac{\partial R}{\partial t} + \sum_{i=2,3} \bigg\{ \braket{U_i} \frac{\partial R}{\partial x_i} - \frac{\partial}{\partial x_i} \left[ \left( \gamma + \gamma_\text{SGS} \right) \frac{\partial R}{\partial x_i}\right] \bigg\}-\\ 
&\Omega_\text{mix}\frac{\partial}{\partial \varphi_2}\left[ (\varphi_2 -\braket{\Phi_2} ) R\right] -C_r \varphi_1 \frac{\partial}{\partial \varphi_2} [\varphi_2 R] \bigg) = 0.
\end{split}
\end{equation}
Note here that it would be possible to solve for $g,h,R$ separately if not for the nonlinear term $\braket{\Phi_\alpha} = \int \varphi_\alpha f \text{d} \varphi_1 \text{d} \varphi_2$. This term can however be eliminated from the above equation by setting $C_\Omega = 0 \rightarrow \Omega_\text{mix} = 0$, thus completely decoupling the solutions along $x_1, \varphi_1$ from each other and the other dimensions. In fact, it is straightforward to show that setting $C_\Omega = 0$ will decouple \emph{all} the dimensions from each other, turning Eq.~(2) of the Main Text into a separable PDE with product-state solutions. Thus at low $C_\Omega$, there should only be limited correlation between the different dimensions.

\subsubsection*{Empirical computational cost}
The Main Text outlines the theoretical complexity of our algorithm. This section shows the empirical computational cost. The CPU-times referenced here are all measured on a \emph{single} CPU core, by selecting the wall-time of the longest of the first $16$ time-steps (out of the full $M_t$ required to complete the time-evolution) of the algorithm (at $C_\Omega=1$, $\text{Da}=0$).

\begin{figure}\begin{center}
\includegraphics[width=.7\linewidth]{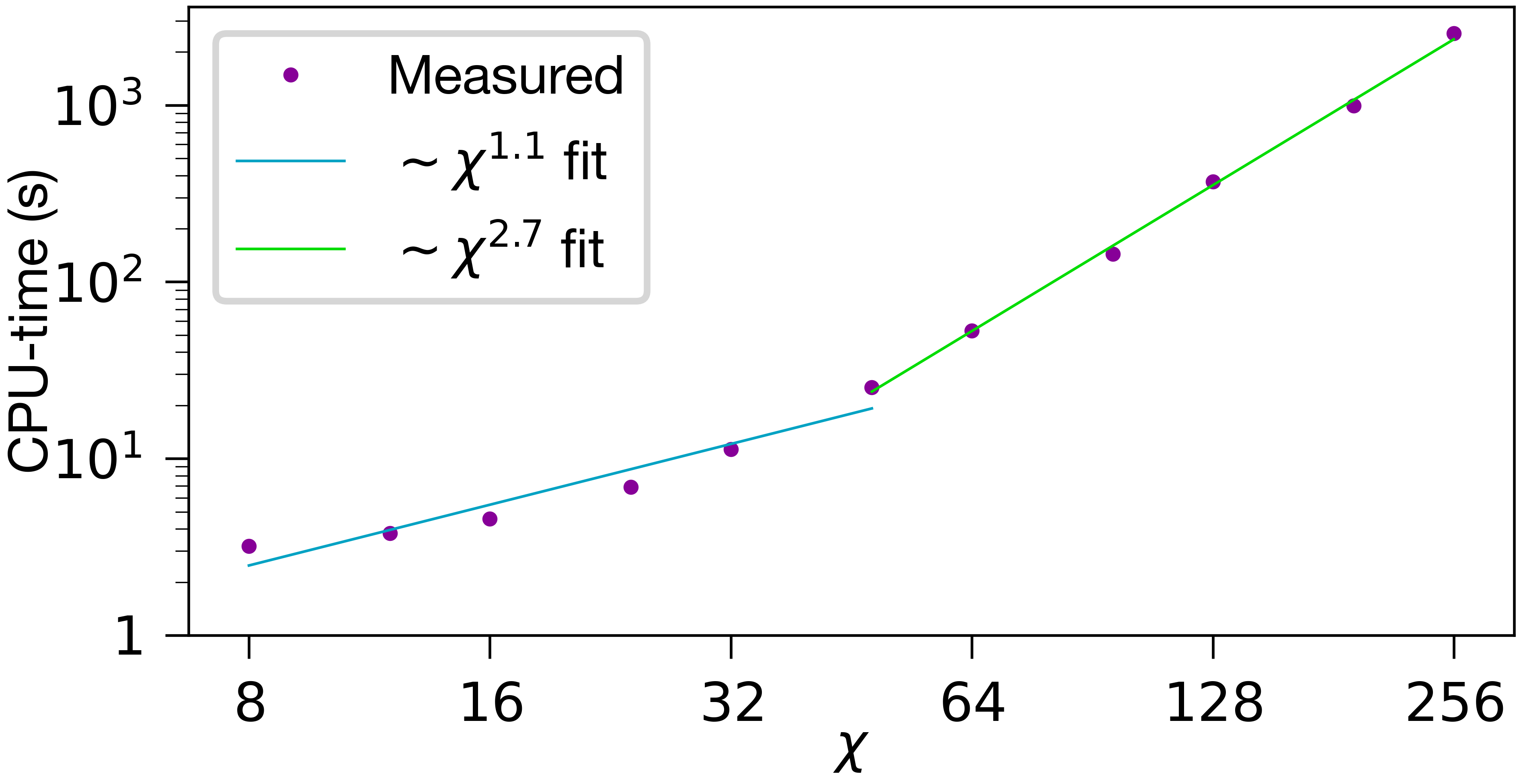}
\caption{\label{fig:cpu_chi}
\textbf{Empirical cost of MPS-TN algorithm}. The single-core CPU-time (in seconds) required for a typical time-step of the MPS algorithm is shown for different $\chi$. The algorithm was implemented in Python 3 [on top of the quimb library~\cite{Gray2018}] and run on a \emph{single} core from a Intel Xeon 8268 CPU.}
\end{center}
\end{figure}

\begin{figure}\begin{center}
\includegraphics[width=.8\linewidth]{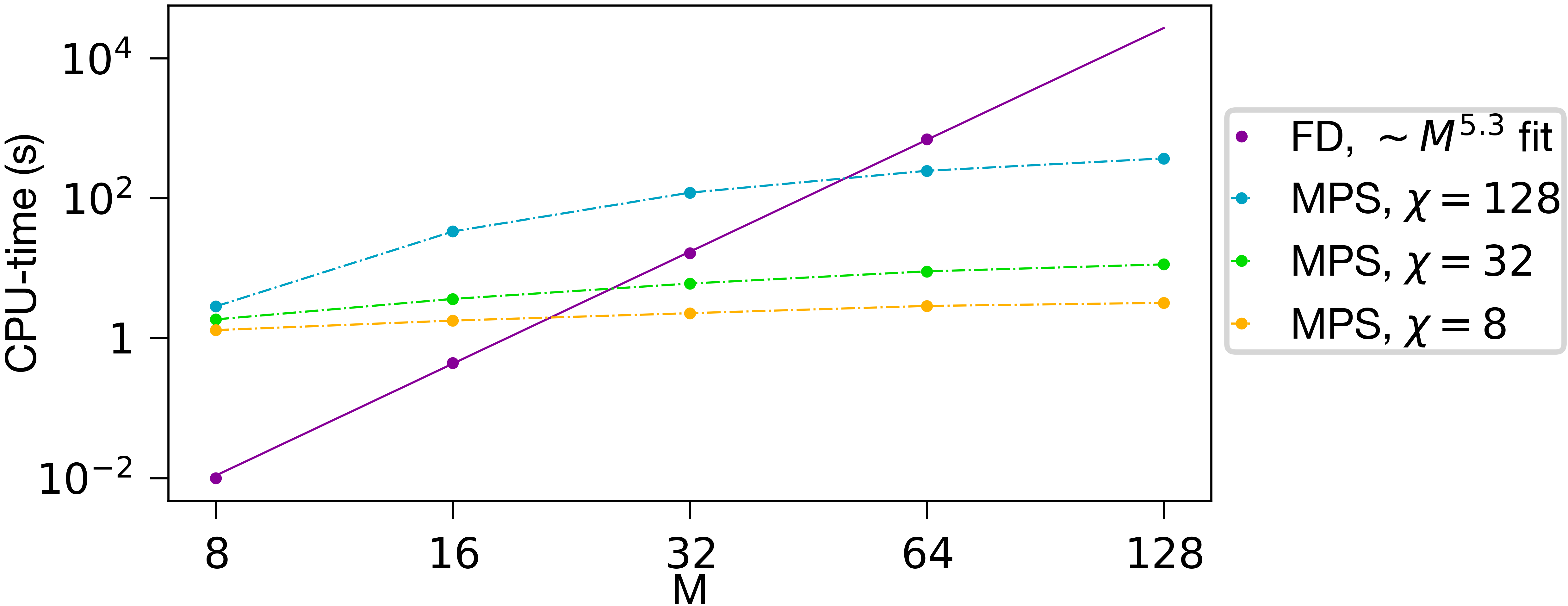}
\caption{\label{fig:cpu_M}
\textbf{MPS-TN versus FD scheme empirical cost}. Equivalent single-core CPU-time to Fig.~\ref{fig:cpu_chi}, except now plotted across $M$ for both the MPS algorithm at various $\chi$ and an equivalent, standard FD scheme (implemented in Python 3). For the FD scheme, the $M=128$ datapoint could not be measured due to the extreme memory load required.}
\end{center}
\end{figure}

Single-core times as functions of $\chi,M$ are shown in Figs.~\ref{fig:cpu_chi} and~\ref{fig:cpu_M}. The first figure shows that at small $\chi$, the cost of the program is dominated by the overhead of the Python interpreter, while at intermediate $\chi$, the $\sim\chi^4$ element-wise products are too few to practically dominate the many $\sim\chi^2$ and $\sim\chi^3$ operations (outlined in the Main Text) of the algorithm. Fig.~\ref{fig:cpu_M} empirically confirms the $\sim \log M$ scaling of the MPS algorithm, in contrast to the $\sim M^d$ cost required by an equivalent, standard FD scheme. Note how the MPS algorithm at $\chi=32, M=128$ requires $\mathcal{O}(1/10^3)$ of the flops of the equivalent FD scheme.

%%%%%%%%%%%%%%%% SUPPLEMENTARY TEXT %%%%%%%%%%%%%%%

% If your supplement is very short you might need to uncomment the following line to avoid
% layout problems with the figures and tables.
%\newpage

%%%%%%%%%%%%%%%% SUPPLEMENTARY FIGURES %%%%%%%%%%%%%%%

Main Text Figs.~2b,~2c can be combined with Fig.~\ref{fig:cpu_chi} to show how the observed accuracy of the MPS algorithm increases with computational effort. This relationship is illustrated in Supplementary Fig.~\ref{fig:err_cpu}. 

\begin{figure}\begin{center}
\includegraphics[width=1.0\linewidth]{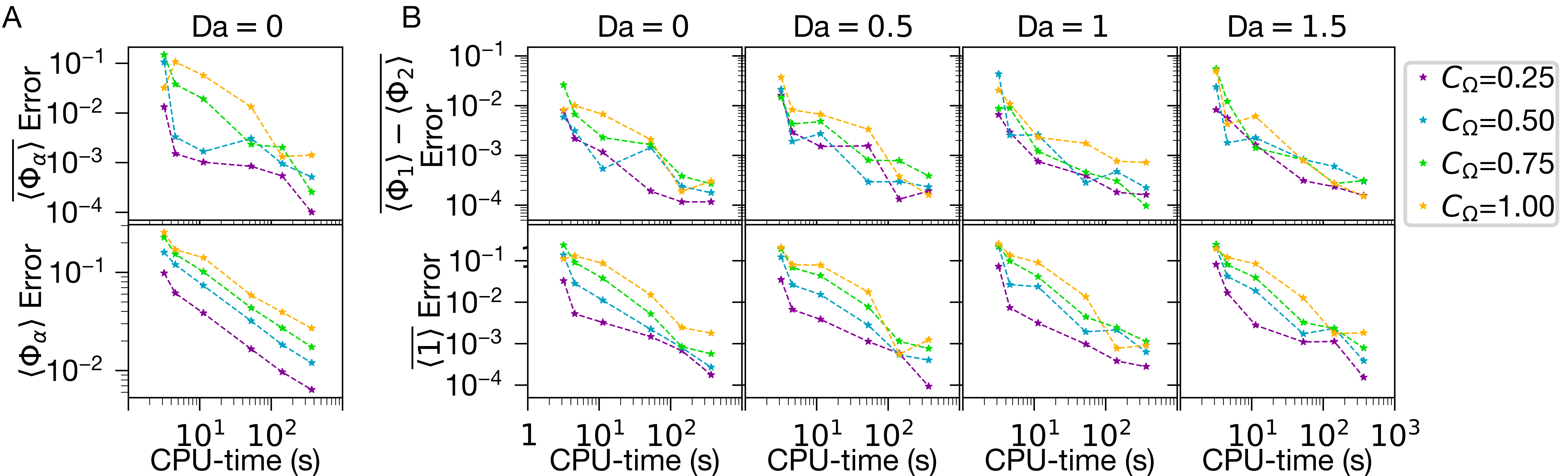}
\caption{\label{fig:err_cpu}
\textbf{Accuracy convergence of MPS-TN algorithm versus empirical computational cost}. Here the errors of Main Text Fig.~2 are plotted as functions of the single-core CPU time measured in Fig.~\ref{fig:cpu_chi}. \textbf{A} corresponds to the Fig.~2b errors, and \textbf{B} to the Fig.~2c ones.}
\end{center}
\end{figure}

%%%%%%%%%%%%%%%% SUPPLEMENTARY TABLES %%%%%%%%%%%%%%%

%%%%%%%%%%% CAPTIONS FOR OTHER SUPPLEMENTARY FILES %%%%%%%%%%

\clearpage % Clear all remaining figures and tables then start a new page

%%%%%%%%%%%%%%%% SUPPLEMENTARY REFERENCES %%%%%%%%%%%%%%%

% Do NOT include a reference list in the supplement.
% All references must be in a single list at the end of the main text.
% The copyeditors will ensure that the correct reference list appears with each version of the paper
% (print, HTML, PDF, mobile app, metadata for bibliographic databases etc.)

\end{document}